\documentclass{article}
\usepackage{spconf,amsmath,graphicx,amssymb}
\usepackage{cite}
\usepackage{algorithm}
\usepackage{algpseudocode}
\usepackage[normalem]{ulem}
\usepackage{color}

\newcommand*{\affaddr}[1]{#1}
\newcommand*{\affmark}[1][*]{\textsuperscript{#1}}

\newcommand\blfootnote[1]{%
  \begingroup
  \renewcommand\thefootnote{}\footnote{#1}%
  \addtocounter{footnote}{-1}%
  \endgroup
}

\title{X-Ray spectral estimation using Dictionary Learning}

\name{\begin{tabular}{c}
  Wenrui Li\affmark[1], Venkatesh Sridhar\affmark[2], K. Aditya Mohan\affmark[2], Saransh Singh\affmark[2],\\ Jean-Baptiste Forien\affmark[2], Xin Liu\affmark[2], Gregery T. Buzzard\affmark[1], Charles A. Bouman\affmark[1]
\end{tabular}}
\address{\affaddr{\affmark[1]Purdue University, West Lafayette, IN 47907, USA}\\
\affaddr{\affmark[2]Lawrence Livermore National Laboratory, Livermore, CA 94550, USA}\\}
\begin{document}
%
\maketitle

\begin{abstract}


As computational tools for X-ray computed tomography (CT) become more quantitatively accurate, knowledge of the source-detector spectral response is critical for quantitative system-independent reconstruction and material characterization capabilities.
Directly measuring the spectral response of a CT system is hard, which motivates spectral estimation using transmission data obtained from a collection of known homogeneous objects.  
However, the associated inverse problem is ill-conditioned, making accurate estimation of the spectrum challenging, particularly in the absence of a close initial guess.

In this paper, we describe a dictionary-based spectral estimation method that yields accurate results without the need for any initial estimate of the spectral response.
Our method utilizes a MAP estimation framework that combines a physics-based forward model along with an $L_0$ sparsity constraint and a simplex constraint on the dictionary coefficients.
Our method uses a greedy support selection method and a new pair-wise iterated coordinate descent method to compute the above estimate.
We demonstrate that our dictionary-based method outperforms a state-of-the-art method as shown in a cross-validation experiment on four real datasets collected at beamline 8.3.2 of the Advanced Light Source (ALS).\blfootnote{Document Release Number: LLNL-CONF-845171}
\end{abstract}


\section{Introduction}
\label{sec:intro}

Non-destructive evaluation (NDE) is an increasingly important application of CT, which motivates improved capabilities for quantitative energy-independent reconstruction and precise material characterization.
Such reconstruction techniques utilize physically accurate forward models that account for the polyenergetic nature of the X-ray source radiation and associated physical phenomena such as beam-hardening\cite{jin2015beamhardening}, rather than simpler models based on mono-energetic approximations.
However, an accurate estimate of the CT system's source-detector spectral response is a necessity for the above methods.
For example, the method for reconstructing energy-independent material properties like effective atomic-number and electron density from dual-energy CT scans \cite{busi2019method,azevedo2016system,champley2019method} and the method for tissue characterization in 
\cite{mccollough2015dual, so2021spectral}
require a precise calibration of the source-detector spectral response.

Direct measurement of the spectral response of the whole X-Ray CT system is difficult since the detector response is hard to measure. 
Based on Beer–Lambert's law\cite{lambert1760photometria}, we can use a linear model for transmission measurements of objects with known dimensions and composition to reconstruct the discretized spectral response of a CT scanner, but this yields a highly ill-conditioned system.
Champley et al. \cite{champley2019method} use linear least-squares to do spectral estimation (LSSE) with constraints to enforce non-negativity. However, this method requires an accurate initial guess close to the true spectrum.
Various regularization methods\cite{ruth1997estimation, yan1999modeling} are used to overcome the issues introduced by the ill-conditioned nature of the problem. 
SVD-based algorithms have also been applied to the spectral estimation problem\cite{tominaga1986530, armbruster2004spectrum, leinweber2017x}.
Sidky et al.\cite{osti_20711775} represent the spectrum as a linear combination of B-splines and use EM (Expectation–maximization) to find the solution.
Zhao et al.\cite{zhao2014indirect} estimate the spectrum as a linear combination of six Monte Carlo model spectra.
Liu et al.\cite{liu20201495} introduce compressed sensing to estimate the spectrum.
The central ideas of the above methods have a common theme: try to solve the ill-conditioned inverse problem either with regularized optimization or by introducing basis spectra to perform the estimation.

In this paper, we introduce a novel dictionary-based spectral estimation (DictSE) method that can efficiently reconstruct the overall spectral response of a CT system from transmission scans of multiple known objects, without the need for accurate initialization.
We represent the unknown spectral response using an over-complete dictionary that accounts for vast combinations of different source spectra, filter attenuation characteristics, and detector energy-response models.
We formulate the reconstruction problem as a MAP estimation framework that combines a linear beam-hardening forward model along with prior constraints.
Specifically, we impose an $L_0$ sparsity constraint to limit the support for the spectrum representation and a simplex constraint to account for the bright-dark normalization of the transmission data.
We present a novel iterative optimization strategy that alternates between support selection and pairwise iterative coordinate descent (ICD) update to find the optimal sparse representation of the spectrum. 
Finally, we demonstrate DictSE through a cross-validation experiment on four datasets collected at beamline 8.3.2 of ALS.

\section{Reconstruction model}
\label{sec:rec_model}
In this section, we describe the spectral estimation problem and our proposed solution in more detail.  We use a physics-based model for X-ray transmission measurements and discretize the model in energy and by projection to produce a linear measurement model.
Then we introduce a dictionary-based framework and express the reconstruction as a MAP estimation problem.
Finally, we solve this dictionary-based MAP problem using support selection to enforce sparse coding and an ICD algorithm modified to enforce a simplex constraint on the coefficients of the selected dictionary elements.

\vspace{-2ex}
\subsection{Physics-based Model and Discretization}
\label{ssec:phy_model}

\begin{figure}[!t]
\centering
\centerline{\includegraphics[width=7cm]{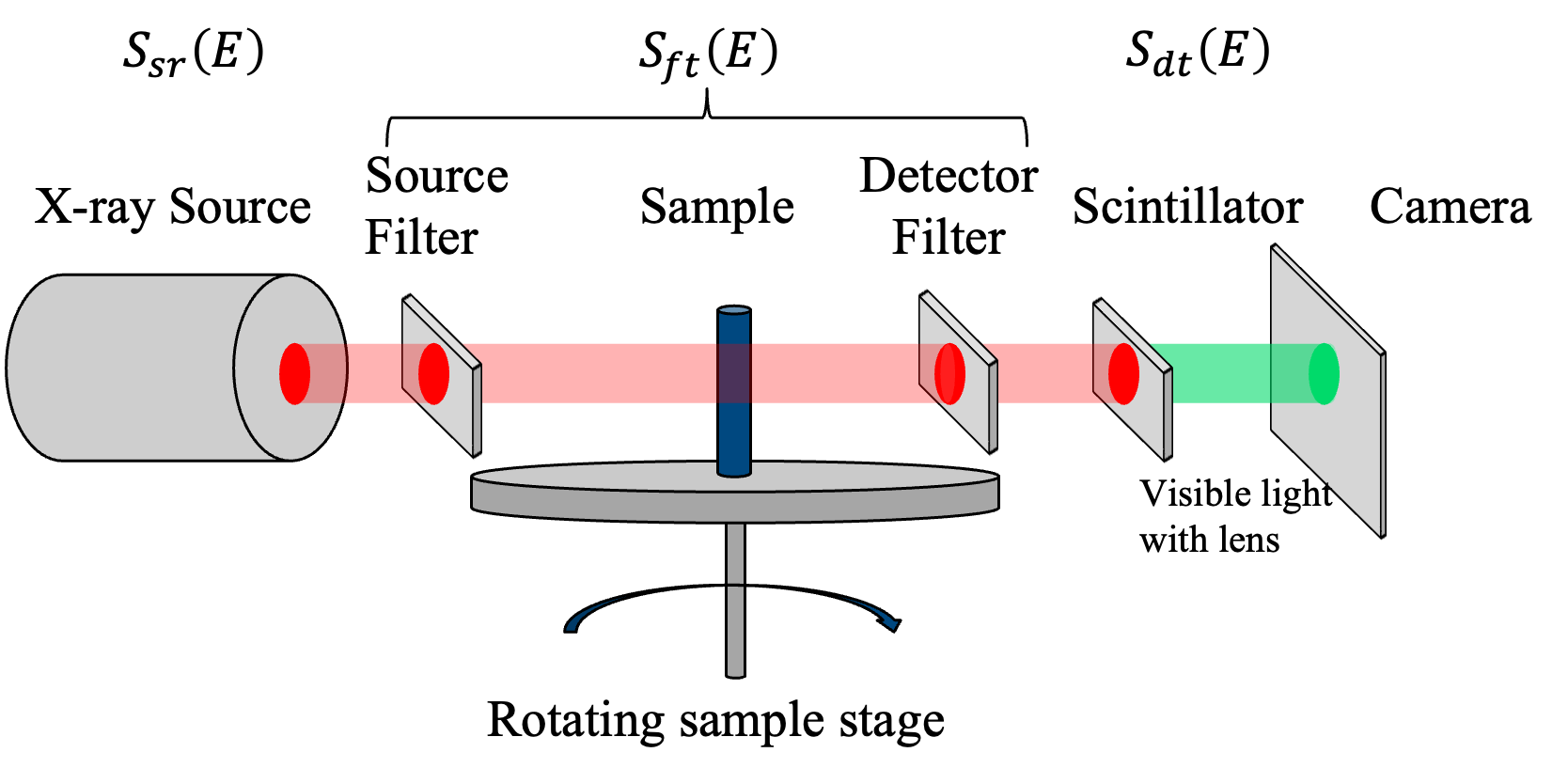}}
\caption{\textbf{Physics model of CT scanning.} A homogenous cylindrical object is scanned. The entire spectral response can be decomposed into the product of the X-ray source spectrum, filter response, and detector response, respectively.}
\label{fig:phy_model}

\end{figure}
Fig.~\ref{fig:phy_model} illustrates the setup for X-ray spectral estimation. 
Using notation as in that figure, the spectral response of this CT system is the product of the X-ray source spectrum, filter response, and detector response, which yields the response function
\begin{equation}
  \label{equ:spec_model}
  S(E)=S_{sr}(E) S_{ft}(E) S_{dt}(E) \ ,
\end{equation}
and our goal is to estimate $S(E)$.

We relate $S(E)$ to measurements by first defining $I$ and $I_0$ to be the measured intensity of the object scan (with sample) and the blank scan (without sample), respectively. 
Based on Beer–Lambert's law, we have
\begin{equation}
  \label{equ:obj_scan}
  I=\int_0^{E_{\max }} S(E) \cdot \exp \left\{-\int_L \mu(E, r) d r\right\} d E \ ,
\end{equation}
where $\mu(E, r)$ is linear attenuation coefficient (LAC) with units $mm^{-1}$.  Likewise, $I_0$ is defined as in \eqref{equ:obj_scan} with $\mu = 0$.  
The (normalized) transmission through line $L$ is then defined as
\begin{equation}
  \label{equ:transm}
  \begin{aligned}
  y & =\frac{I}{I_0}=\int_0^{E_{\max }} \bar{S}(E) \cdot \exp \left\{-\int_L \mu(E, r) d r\right\} d E \ ,
  \end{aligned}
\end{equation}
where $\bar{S}(E)=\frac{S(E)}{\int_0^{E_{\max }} S(E) d E}$ is the normalized response. 

We assume each sample consists of solid rods made from known materials taken from a specified reference set $\Phi$.
For each material $s\in \Phi$ and projection $i\in\{0,...,M-1\}$, we define $L_{i, s}$ to be the path length of the $i^{th}$ projection through the $s^{th}$ material and $\mu_s(E)$ to be the LAC of the $s$  material at energy $E$.  
Assuming that $I_0$ is independent of projection and noise-free, this yields the transmission for the $i^{th}$ projection as 
\begin{equation}
  \label{equ:scaled_transm}
  \begin{aligned}
  y_i & =\int_0^{E_{\max }} \bar{S}(E) \cdot \exp \left\{-\sum_{s \in \Phi} \mu_s(E) L_{i, s}\right\} d E+\tau_i \ ,
  \end{aligned}
\end{equation}
where $\tau_i$ is additive noise.

We discretize in energy by subdividing into non-overlap\-ping bins $[E_j, E_{j+1}]$ and defining $x_j=\int_{E_j}^{E_{j+1}} \bar{S}\left(E_j\right) d E$.  
Making the approximation that each $\mu_s$ is constant on each bin, we define the coefficient from the $j^{th}$ energy bin to the $i^{th}$ projection as $F_{i, j}=\exp \left\{-\sum_{s \in \Phi} \mu_s\left(E_j\right) L_{i, s}\right\}. $
Then \eqref{equ:scaled_transm} becomes
\begin{equation}
  \label{equ:y_discrete}
  \begin{aligned}
  y_i & \approx \sum_{j=0}^{J-1} \int_{E_j}^{E_{j+1} } \bar{S}(E) F_{i,j} d E+\tau_i = \sum_{j=0}^{J-1} F_{i,j} x_j + \tau_i \ .
  \end{aligned}
\end{equation}

Using this along with constraints to yield a normalized spectrum, the forward model is 
\begin{equation}
  \begin{aligned}
  \label{equ:smp_model}
  &Y=F x + \tau, \\
  &\text{s.t. } \|x\|_1=1, x_j \geq 0 \; \forall j,
  \end{aligned}
\end{equation}
where $Y\in \mathbb{R}^M$ is a vector of $M$ normalized transmission measurements, $F \in \mathbb{R}^{M \times N_e}$ is a forward matrix over $N_e$ energy bins,  $\tau$ is additive noise, and $x\in \mathbb{R}^{N_e}$ is the unknown vector discretization of $\bar{S}\left(E\right)$ that we seek to estimate.

\vspace{-2ex}
\subsection{Dictionary-based Model}
\label{ssec:dict_model}
Developing a dictionary-based method instead of directly estimating the spectrum at each energy bin has several motivations.
When the discretization of $x$ is very fine, the projection model matrix $F$ has a large null space, leading to an underdetermined reconstruction, which requires a good initial estimate.
Also, with a dictionary-based method, we can apply a sparsity-promoting penalty so that the dimensionality of the optimization problem can be significantly decreased and the time for reconstruction reduced.

More specifically, as indicated in \eqref{equ:smp_model}, we model $x$ as a discrete probability distribution, so that the entries of $x$ are nonnegative and sum to 1. We call this the simplex constraint and use $\mathcal{S}$ to represent the set of all such possible vectors.

We use a fixed, over-complete dictionary $D$ to represent the unknown normalized spectrum, $x$, as
\vspace{-1ex}
\begin{equation}
  x=D\omega
\end{equation}
where $D$ is an $\mathbb{R}^{N_e\times N_k}$ matrix, each column $D_{*,k}\in \mathcal{S}$ represents a normalized basic spectrum.
With this $D$, the transmission model can be rewritten as 
\vspace{-1ex}   
\begin{equation}
  \label{equ:dict_model}
  Y=F D \omega + \tau \ .
\end{equation}

\vspace{-2ex}
\subsection{MAP Estimate}
\label{ssec:map}
We use a Bayesian framework to estimate $\omega$ from transmission measurements under sparsity and simplex constraints.  We define transmission weights using a diagonal matrix $\Lambda$, where we take $\Lambda_{i,i} = 1/(y_i M)$.  We then define 
the loss function $l(\omega) = \frac{1}{2}\|Y-F D \omega\|_{\Lambda}^2$, in which case the MAP estimate is given by  
\begin{equation}
  \label{equ:map_model}
  \widehat{\omega}=\arg \min _{\substack{\omega \in \mathcal{S} \\\|\omega\|_0 \leq C}}l(\omega) \ , 
\end{equation}
where $\omega \in \mathcal{S}$ enforces $x \in \mathcal{S}$ since each column of $D$ is a simplex;  the $L_0$ constraint ensures $\omega$ has no more than $C$ non-zero components. 

\begin{algorithm}[!t]
  \caption{Dictionary-based Spectrum Reconstruction Algorithm (see equation \eqref{equ:kbfw_model} for notation)} 
  \label{alg:mp_est}
  \begin{algorithmic}[1]
  
  \State{Initialize $k^* \gets \arg \min _{\substack{k }}l(\epsilon_{k})$,  $\omega=\epsilon_{k^*}$, $\Omega=\{k^*\}$} 
  \While{$|\Omega| < C$}  \mbox{\hspace{12pt}}  // Support selection
  \State{$\left(k^*, \beta^*\right) \gets \arg \min _{\substack{k \notin \Omega \\ \beta \in[0,1)}}l'(k,\beta | \widehat{\omega})$ } 
  \State{$\omega \gets \beta^* \omega+\left(1-\beta^*\right) \epsilon_{k^*}$}
  \State{$\Omega \gets \Omega \cup \{k^*\}$}
  \State{$g \gets k^*$}  \mbox{\hspace{12pt}} // Select new element for next loop
      \While{not converged}  \mbox{\hspace{12pt}} // Pairwise ICD update
          \For{$k\in \Omega$ and $k\neq g$}
          \State{$\alpha^* \leftarrow \arg \min _{\alpha \in\left[-\omega_k, \omega_g\right]}l\left(\omega+\alpha (\epsilon_k -\epsilon_g)\right)$}
          \State{$\omega \leftarrow \omega+\alpha^* (\epsilon_k -\epsilon_g)$}
          \EndFor   
      \EndWhile
  \EndWhile
  \State{$\hat{\omega} \gets \omega$ \mbox{\hspace{12pt}}  // Return the final estimate }
  \end{algorithmic}
\end{algorithm}

\vspace{-2ex}
\subsection{Support Selection and Pairwise ICD update}
\label{ssec:ds_pwicd}
To minimize the MAP cost function, we alternate between greedy support selection and a pairwise ICD update.
Inspired by the Orthogonal Matching Pursuit (OMP)\cite{pati1993}, our DictSE builds support set $\Omega=\{k: \omega_k \neq 0\}$ by adding one basic spectrum from the dictionary at a time and then updating the coefficients.
However, the conventional support selection method in OMP does not account for  simplex constraints on $\omega$.
Further, the OMP method assumes that the dictionary atoms are normalized, whereas, in our spectral estimation problem, the product of forward matrix $F$ and the spectrum dictionary $D$ is not. 
Thus choosing a basis spectrum with conventional matching pursuit is inappropriate.

To describe our alternative method for support selection, we first define a function $l'(k,\beta | \widehat{\omega})$ that measures the fit to data obtained by scaling the existing coefficients $\hat{\omega}$ by $\beta$ and using the remaining weight on the $k^{th}$ coefficient.  That is, 
\begin{equation}
  \label{equ:kbfw_model}
  \begin{aligned}
  l'(k,\beta | \widehat{\omega})&=l(\beta \widehat{\omega}+\left(1-\beta\right) \epsilon_{k})\\
  &=\frac{1}{2}\left\|Y-\beta F D \widehat{\omega}-\left(1-\beta\right) F D_{*, k}\right\|_{\Lambda}^2
  \end{aligned}
\end{equation}
where $\beta \in [0,1)$ enforces $\omega \in \mathcal{S}$ and $\epsilon_{k}$ is a one-hot vector for the $k^{th}$ spectrum.

Then we select a new element from the dictionary by  minimizing $l'(k,\beta | \widehat{\omega})$ to obtain
\vspace{-1ex}
\begin{equation}
  \begin{aligned}
  \left(k^*, \beta^*\right) &\leftarrow 
  \arg \min _{\substack{k \notin \Omega\\ \beta \in[0,1)}}\left\{l'(k,\beta | \widehat{\omega})\right\}.
\end{aligned}
\end{equation}
This minimization can be solved easily for each $k \notin \Omega$ since equation (\ref{equ:kbfw_model}) is quadratic in the scalar $\beta$. In fact, defining $e_k = FD\widehat{\omega}- FD_{*,k}$, we have 
\vspace{-1ex}
\begin{equation}
\beta_k=\frac{e_k^T \Lambda (Y-FD_{*,k})}{\| e_k\|^2_{\Lambda}} \ .
\end{equation}
Using $\beta_k$ in equation \eqref{equ:kbfw_model} allows us to find $k^*$ that minimizes equation \ref{equ:kbfw_model}; this $k^*$ is then included in $\Omega$ with $\omega_{k^*} = 1- \beta^*$ and the remaining $\omega_k$ scaled by $\beta^*$.

To rebalance the weights while enforcing the simplex constraint, we use pairwise ICD between the most recently added dictionary element with index $g$ and the remaining elements in $\Omega$, as shown in lines 6-12 of Algorithm \ref{alg:mp_est}.
More precisely, after choosing $g$, we loop repeatedly over  $k \in \Omega \setminus \{g\}$, in each case finding an optimal pairwise update $\omega \leftarrow \omega+\alpha^* (\epsilon_k -\epsilon_g)$, where $\alpha^*$ is chosen by
\vspace{-1ex}
\begin{equation}
  \alpha^* \leftarrow \arg \min _{\alpha \in\left[-\omega_k, \omega_g\right]}\left\{l\left(\omega+\alpha (\epsilon_k-\epsilon_g) \right)\right\} \ .
\end{equation}
The constraints on $\alpha$ in the minimization ensure each $\omega_k \geq 0$. Since $l(\omega)$ is quadratic, $\alpha^*$ can be computed as below
\vspace{-1ex}
\begin{equation}
  \alpha^*=\text{Clip}\left\{\frac{(Y-F D \omega)^{\mathrm{T}} \Lambda F D (\epsilon_k -\epsilon_g)}{\left\|F D (\epsilon_k -\epsilon_g)\right\|_{\Lambda}^2}, [-\omega_k, \omega_g]\right\}
\end{equation}
The pairwise ICD update will stop when the total update is less than $10^{-6}$.
The algorithm is summarized in Algorithm \ref{alg:mp_est}.

\begin{table}[t]
  \centering
  \caption{Setup for X-ray Scanning}
  \begin{tabular}{ll}
    \hline
    Projection Geometry: & Parallel beam geometry \\ 
    Source filter: & 2 $mm$ Silicon \\
    Scintillator: & 50 $\mu m$ $Lu_{3}Al_{5}O_{12}$ \\
    Max Energy: & 100 KeV \\ 
    Views Spanning: & Equi-spaced in $[0,2\pi]$ \\
    Detector Pixel Size: & 0.00065 mm \\
    Nviews $\times$ Nrows $\times$ Ncolumns: & $2625 \times 100 \times 2560$ \\ 
    Sample-detector distance: & $0.3$ mm \\ \hline
  \end{tabular}
\label{table:setup_dataset}
\end{table}

\label{sec:imp}
\begin{figure}[!t]
    \centering
    \centerline{\includegraphics[width=8.5cm]{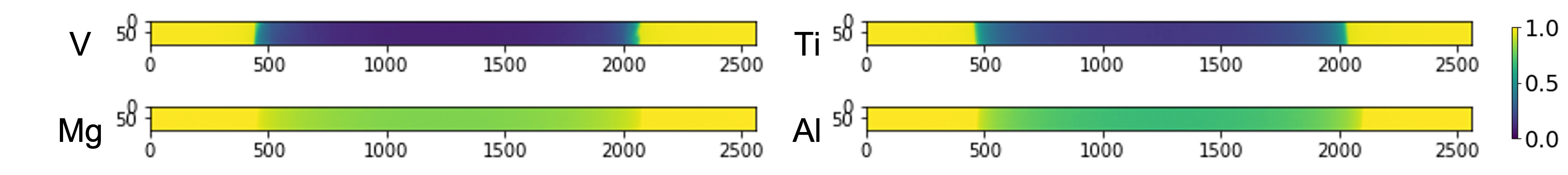}}
  \caption{\textbf{Real scans of 4 rods with different materials.}}
  \label{fig:scans}
\end{figure}

\section{ Implementation}
\label{sec:implementation}
  
In this section, we describe the normalized transmission data $Y$, the projection matrix $F$, and the dictionary of spectra $D$, which are required to estimate the spectral response of an X-ray CT system using Algorithm \ref{alg:mp_est}.

\vspace{-2ex}
\subsection{ Transmission data $Y$}
\label{ssec:trans_data}

We collected four CT datasets of different metal rods, $\Phi = \{Ti, V, Al, Mg\}$, at beamline 8.3.2 of the ALS. 
As shown in Fig.~\ref{fig:scans}, for each dataset, we scanned a single rod.
The same CT scanner setup was used to collect all datasets.
Table~\ref{table:setup_dataset} gives more detailed information about the X-ray CT measurements.

We scanned $15$ bright scans and $10$ dark scans and averaged them to obtain $\bar{I}_{\text {bright }}$ and $\bar{I}_{\text {dark }}$. 
Then, for each view, we normalized measurement data $I_{\text {scan }}$ to obtain
$  Y=\frac{I_{\text {scan }}-\bar{I}_{\text {dark }}}{\bar{I}_{\text {bright }}-\bar{I}_{\text {dark }}}$.

\vspace{-2ex}
\subsection{ Forward Matrix $F$}
\label{ssec:fw_mat}
For each dataset, the sample is a single rod that is solid and pure. 
To calculate $F_{i,j}$, we need the path length of the $i^{th}$ projection through this rod.  
To obtain the path length, we used filtered back projection to reconstruct the volume of the rod and then generated a mask to represent the object area. 
Using this mask, we calculated the path length for each projection and used this to calculate the forward matrix $F$.

\vspace{-2ex}
\subsection{ Dictionary Generation $D$}
\label{ssec:dict_gen}
As mentioned in equation (\ref{equ:spec_model}), the spectral response can be modeled as the product of the X-ray source spectrum, filter response, and detector response, so we can create a dictionary by varying one or more of these elements.
The X-ray source spectrum is fixed in this experiment; we used an estimate provided by the beamline scientists at the ALS Beamline 8.3.2.
Therefore, in this implementation, we generated the dictionary by varying the filter and detector responses.
Based on the spectrum models in Ref.~\cite{champley2019method}, the filter response and detector response are determined by their material properties and thicknesses, which we vary as in Table~\ref{table:dict} to generate our dictionary.
By combining two groups of responses, we obtain an over-complete dictionary $D$ containing $60\times36=2160$ normalized responses spectra.

\begin{table}[t]
  \centering
  \caption{Dictionary Generation List}
  \begin{tabular}{|cccc|}
  \hline
  \multicolumn{4}{|c|}{Filter response}                                                                                                                                                                                                                         \\ \hline
  \multicolumn{1}{|c|}{Material}  & \multicolumn{1}{c|}{\begin{tabular}[c]{@{}c@{}}Thickness Range\\ $mm$\end{tabular}} & \multicolumn{1}{c|}{\begin{tabular}[c]{@{}c@{}}Step\\ $mm$\end{tabular}} & \begin{tabular}[c]{@{}c@{}}\# of \\ responses\end{tabular} \\ \hline
  \multicolumn{1}{|c|}{$Al$}        & \multicolumn{1}{c|}{0.1$\sim$5.9}                                                   & \multicolumn{1}{c|}{0.2}                                                 & 30                                                         \\ \hline
  \multicolumn{1}{|c|}{$Cu$}        & \multicolumn{1}{c|}{0.2$\sim$0.49}                                                 & \multicolumn{1}{c|}{0.01}                                               & 30                                                         \\ \hline
  \multicolumn{4}{|c|}{Detector response(Scintillator)}                                                                                                                                                                                                                       \\ \hline
  \multicolumn{1}{|c|}{$Lu_{3}Al_{5}O_{12}$} & \multicolumn{1}{c|}{0.025$\sim$0.095}                                               & \multicolumn{1}{c|}{0.002}                                               & 36                                                         \\ \hline
  \end{tabular}
\label{table:dict}
\end{table}
\section{ Experimental Results}
\label{sec:exp}
We compared our proposed DictSE method with a least-squares spectral estimation (LSSE) method provided by Livermore tomography tools (LTT) \cite{CHAMPLEY2022102595} on four datasets described in section \ref{ssec:trans_data}. 
An initial spectrum for the LSSE method was generated by LTT using $3 mm$ silicon as a source filter and $50 \mu m$ $Lu_{3}Al_{5}O_{12}$ as a scintillator.
We then evaluated the estimated spectra of DictSE and LSSE using leave-one-out cross-validation since we do not have a ground truth response.


Table~\ref{table:res} demonstrates that DictSE's reconstructed spectra outperform the LSSE's reconstructed spectra in NRMSE for all cross-validation cases.
For each case $v\in\{1,2,3,4\}$, we computed NRMSE $= \frac{\| Y_{v}-\widehat{Y}_{v}\|_{2} }{\| Y_{v}\|_{2} }$  to compare transmission measurements $Y_{v}$ and transmission value of the forward model using estimated spectrum $\widehat{Y}_{v}=F_{v}D\widehat{\omega}_v$ on the validation rod.

\begin{figure}[!t]
  \centering
  \centerline{\includegraphics[width=8.5cm]{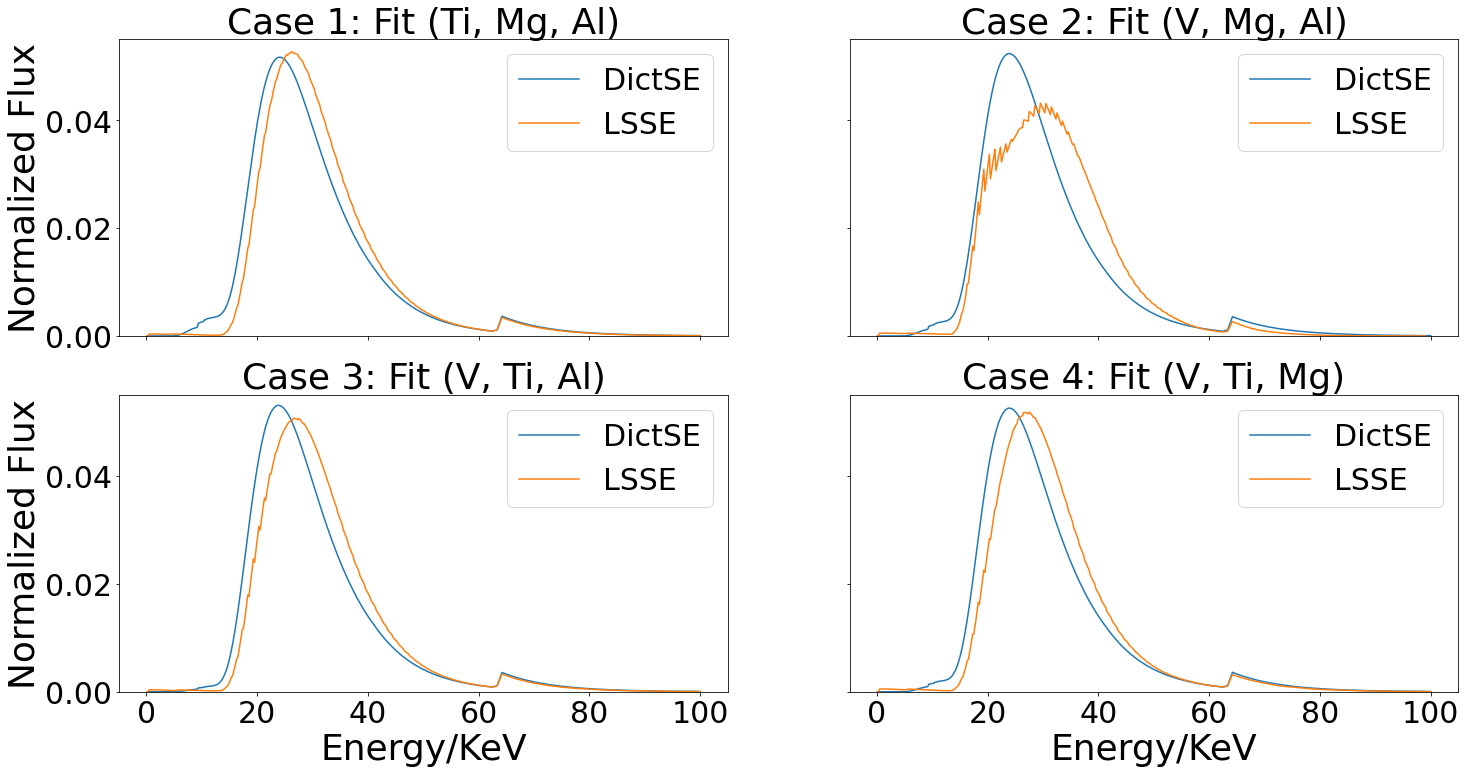}}
  \caption{\textbf{Reconstructed Spectra with DictSE and LSSE.} }
  \label{fig:est_sp}
  \end{figure}

Fig.~\ref{fig:est_sp} shows all four cases of cross-validation reconstructed spectra using both the DictSE and LSSE methods. 
For each case, DictSE's estimated spectra are smoother than the LSSE's.
Also, from the shape of the reconstructed spectra over all cases, DictSE is less data-sensitive than LSSE.
\begin{table}[!t]
  \centering
  \caption{Leave-One-Out Cross-Validation NRMSE}
  \begin{tabular}{c|c|c|c|c}
    \hline
  Case & Fit & Test & LSSE & DictSE \\
  \hline
  1 &  $Ti, Mg, Al$ & $V$   & 0.0331    & \textbf{0.0315}   \\
  2 &  $V, Mg, Al$ & $Ti$   & 0.0624    & \textbf{0.0242}   \\
  3 &  $V, Ti, Al$ & $Mg$   & 0.0343    & \textbf{0.0122}   \\
  4 &  $V, Ti, Mg$ & $Al$   & 0.0483    & \textbf{0.0093}   \\
  \hline
\end{tabular}
  \label{table:res}
\end{table}

\section{ Conclusion}
\label{sec:conclude}
This work provides a novel application of dictionary learning to X-Ray spectral estimation, allowing an efficient spectrum reconstruction from a vast dictionary obtained from CT datasets. 
Our method uses a greedy support selection method to do sparse coding followed by pairwise ICD to do minimization while enforcing a simplex constraint.
Leave-one-out cross-validation experiments on four datasets demonstrated that our DictSE method outperforms the LSSE method in NRMSE.

\section{ACKNOWLEDGMENTS}
This work was performed under the auspices of the U.S. Department of Energy by Lawrence Livermore National Laboratory under Contract DE-AC52-07NA27344 and LDRD project 22-ERD-011. The authors acknowledge Dula Parkinson for his support during beamtime.  Charles Bouman was partially supported by the Showalter Trust, and Greg Buzzard was partially supported by NSF CCF-1763896.  
\vfill\pagebreak

\bibliographystyle{IEEEbib}
\bibliography{strings,refs}

\end{document}